\documentclass{article}
\usepackage[resetlabels,labeled]{multibib}
\newcites{Supp}{Supplemental References}

\usepackage{arxiv}
\usepackage{multicol}
\usepackage{pifont}

\usepackage[utf8]{inputenc} 
\usepackage[T1]{fontenc}    
\usepackage{hyperref}       
\usepackage{url}            
\usepackage{booktabs}       
\usepackage{amsmath,amsfonts,amssymb,amsthm}       

\usepackage{nicefrac}       
\usepackage{microtype}      
\usepackage{lipsum}
\usepackage[labelformat=empty]{caption}
\usepackage{float}
\newenvironment{Figure}
  {\par\medskip\noindent\minipage{\linewidth}}
  {\endminipage\par\medskip}
\usepackage{graphicx}

\title{Random Embeddings and Linear Regression can Predict Protein Function}

\author{
    Tianyu Lu \\
  Computer Science\\
  Cell and Systems Biology\\
  University of Toronto\\
  \texttt{tianyu.lu@mail.utoronto.ca} \\
   \And
   Alex X. Lu \\
  Computer Science\\
  University of Toronto\\
  \texttt{alexlu@cs.toronto.edu}
  \And
   Alan M. Moses \\
   Computer Science \\
   Cell and Systems Biology \\
   University of Toronto \\
   \texttt{alan.moses@utoronto.ca} \\
}

\begin{document}
\maketitle

\begin{abstract}

Large self-supervised models pretrained on millions of protein sequences have recently gained popularity in generating embeddings of protein sequences for protein function prediction. However, the absence of random baselines makes it difficult to conclude whether pretraining has learned useful information for protein function prediction. Here we show that one-hot encoding and random embeddings, both of which do not require any pretraining, are strong baselines for protein function prediction across 14 diverse sequence-to-function tasks.

\end{abstract}


\begin{multicols}{2}
\section*{Introduction}

Accurate protein function prediction from sequence enables the design of novel proteins with desired properties. Here, we consider protein function to be any measurable property of a protein sequence that falls within a continuous range of values. It can be the brightness of a fluorescent protein, the binding affinity to a target, the turnover number of an enzyme, or the melting temperature \cite{volk2020biosystems}. Thus, our sequence-to-function models are regression models. Protein function with discrete classes such as cellular localization are not considered here \cite{volk2020biosystems}. To build a generalizable model, much recent work has focused on training embedders with self-supervised learning on large protein sequence databases. Embedders transform a protein's discrete, variable-length amino acid sequence into a continuous, fixed-length vector. The goal is to generate embeddings which capture biophysical information of protein sequence useful for downstream tasks such as protein function prediction \cite{rao2019evaluating}. A top model, typically linear regression, is trained using the embeddings as input. Typically, the embedding representation of a sequence is the average over the position-wise embeddings \cite{bioembeddings}. A survey of 39 biological sequence embedding methods is given in \cite{iuchi2021representation}. 

Recently, using one-hot encoding as input to the top model, which requires no pretraining, was shown to achieve similar performance to pretrained embeddings \cite{shanehsazzadeh2020transfer}. Here, we treat one-hot encoding as a baseline for evaluating embedders on protein function prediction and introduce another simple baseline with strong performance: untrained embedders with random weights. We show their remarkable strengths across 14 function prediction tasks, including extrapolation to sequences with unobserved residue variations and possibly epistatic mutations. Without assessing the performance of random embeddings, we cannot conclude whether the computationally expensive pretraining of embedders on large sequence databases is necessary for protein function prediction.

Here, we ask whether random embeddings can predict protein function just as well as pretrained embeddings, and whether some random embedders do better than others. To the best of our knowledge, no work has assessed the performance of random embeddings for protein function prediction. We also include another baseline, one-hot encoding, following work showing its competitive performance \cite{shanehsazzadeh2020transfer}. We use the full one-hot encoding which retains positional information instead of the averaged one-hot encoding used in TAPE \cite{rao2019evaluating}.

\end{multicols}

\begin{table}[h]
    \centering
    \begin{tabular}{ccccc}
    \hline
    Model    & Parameters & Embedding Dimension & Training Data & Training Time \\
    \hline
      ProtBert Pretrained  & 420M & 1024 & 2.1B & 23.5 days on 1024 GPUs \\
      CPCProt Pretrained  & 1.7M & 512 & 32.2M & 14 days on 2 GPUs \\
      Bepler Pretrained  & 31M & 121 & 21.8M & 3 days on 1 GPU \\
      ProtBert Random  & 420M & 1024 & 0 & 0 \\
      CPCProt Random  & 1.7M & 512 & 0 & 0 \\
      Bepler Random  & 31M & 121 & 0 & 0 \\
      One Hot & 0 & $L\times 21$ & 0 & 0\\
    \hline
    &&&&
    \end{tabular}
    \caption{\textbf{Table 1}: Summary of the embedders and baselines used. The dimension of one-hot encoding is the product of the number of amino acids in a sequence, $L$, and the number of dimensions used to represent one amino acid.}
\end{table}

\begin{multicols}{2}

\section*{Related Work}

A random embedding is obtained by feeding a sequence through an untrained embedder with randomly initialized weights. Using random embeddings for prediction tasks goes against conventional thinking that trained embedders capture biophysical properties of proteins in their embeddings, since random embeddings are not expected to capture such properties. However, random embedders are strong baselines in point clouds classification \cite{sanghi2020powerful}, sentiment analysis \cite{wieting2019no}, text summarization \cite{pilault2020impressive}, and GLUE \cite{wang2018glue}, a collection of tasks for evaluating Natural Language Processing (NLP) models \cite{wang2018can}. Beyond NLP, random neural networks can be used for object detection \cite{saxe2011random}, face recognition \cite{baek2019spontaneous}, and audio source separation \cite{chen2019j}. More examples are given in \cite{wieting2019no, random-review}. In many cases, authors report the surprising result that random embedders can achieve similar or better performance compared to a trained embedder. Further, some random embedders do better than others, which works have attributed to differences in model architecture \cite{wallace2019nlp}.

The "worryingly strong" \cite{wang2018can} performance of random baselines can be explained by Cover's Theorem (1965), which says that random projections are more likely to be linearly separable in higher dimensions \cite{cover1965geometrical}. 27 years later, Schmidt \textit{et al.} proposed a random neural network for classification where all hidden units are given random weights and only the output layer is trained \cite{schmidt1992feed}. Results achieved by random embeddings were near Bayes-optimal. Their network is analogous to using random embeddings to predict protein function: only the top model's parameters are trained while the embedding is a random projection of an input sequence.

\section*{Methods}

We assess pretrained ProtBert \cite{elnaggar2020prottrans}, CPCProt \cite{lu2020self}, and Bepler embedders \cite{bepler2019learning}, along with their randomly initialized counterparts. We chose these three embedders because of their differences in architecture, embedding dimension, number of trainable parameters, and training objectives, summarized in \textbf{Table 1}. For any protein sequence $\mathbf{x}$, $\text{Embed}(\mathbf{x}) \in \mathbb{R}^d$ where $d=1024$ for ProtBert, $d=512$ for CPCProt, and $d=121$ for Bepler. We also introduce a generic random embedder called "Random MLP", which is an embedding layer followed by two linear layers with sigmoid activations, again with randomly initialized weights. Each layer has the same dimension. The generic random embedder has none of the architectural priors present in the other three embedders. We use bioembeddings for the pretrained and random embedders \cite{bioembeddings}. The datasets used and train/test split composition are described in the supplemental.

\section*{Results}

\subsection*{Baselines are Strong Across Diverse Tasks}

\begin{Figure}
    \centering
  \includegraphics[width=\linewidth]{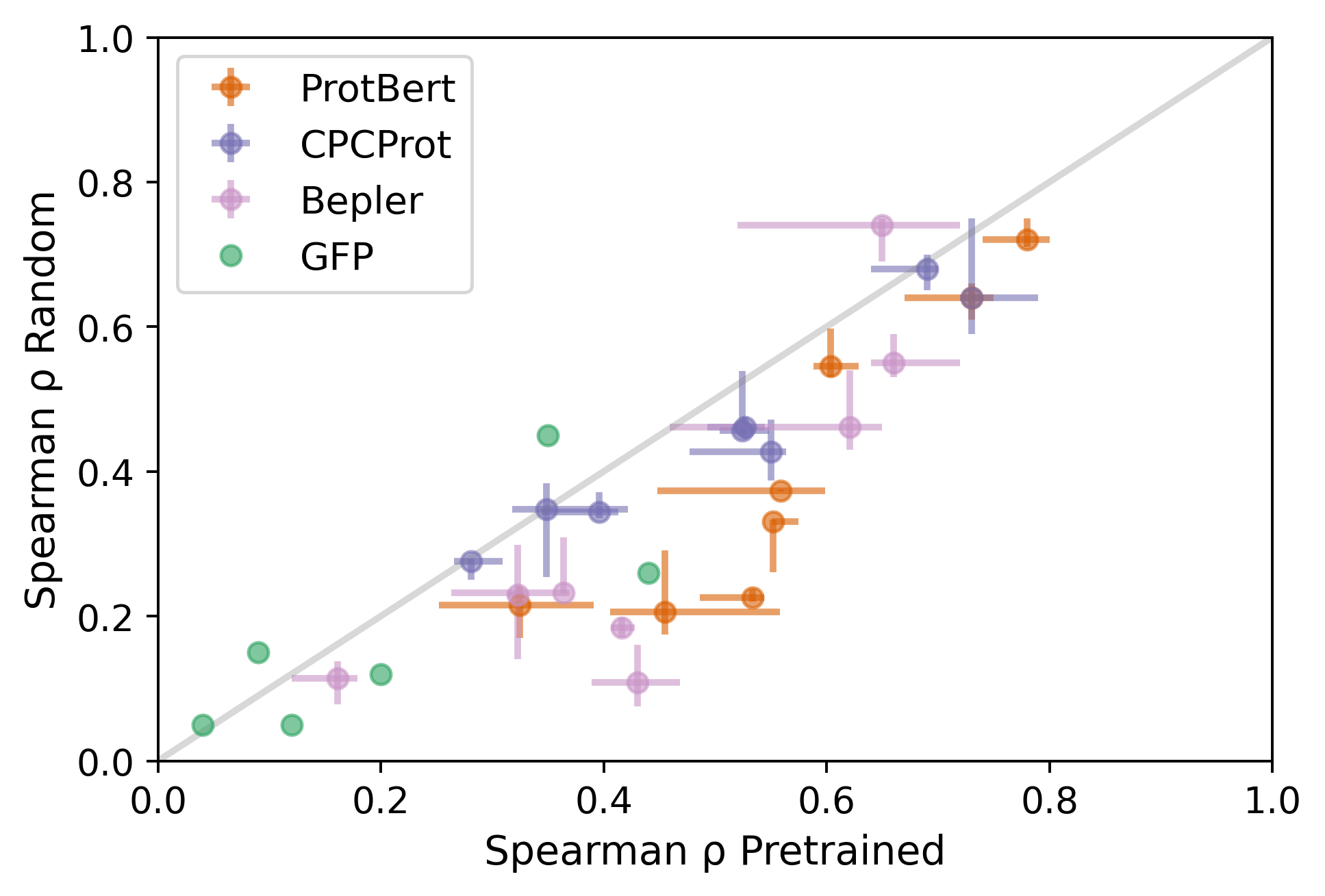}
  \captionof{figure}{\textbf{Figure 1:} Random vs. pretrained embeddings. Error bars cover the max and min value over three replicates, with the median shown as a dot. Points near or above the grey line indicate little to no benefit of pretraining. Performance of all three embedders on the GFP2 and GFP3 splits is denoted as GFP. Dataset details are described in the supplemental.}
\end{Figure}

On six datasets from DeepSequence \cite{riesselman2018deep} and two splits from GFP data \cite{sarkisyan2016local}, we find that randomly initialized embedders can achieve comparable performance to pretrained embedders (\textbf{Figure 1}). In two fluorescence prediction tasks constructed from the GFP dataset \cite{sarkisyan2016local}, we find that the one-hot encoding baseline consistently surpasses the performance of all pretrained embedders assessed (\textbf{Figure 2}). 

\end{multicols}
\begin{figure*}[h]
    \centering
    \hspace*{-0.8cm}
  \includegraphics[width=0.86\textwidth]{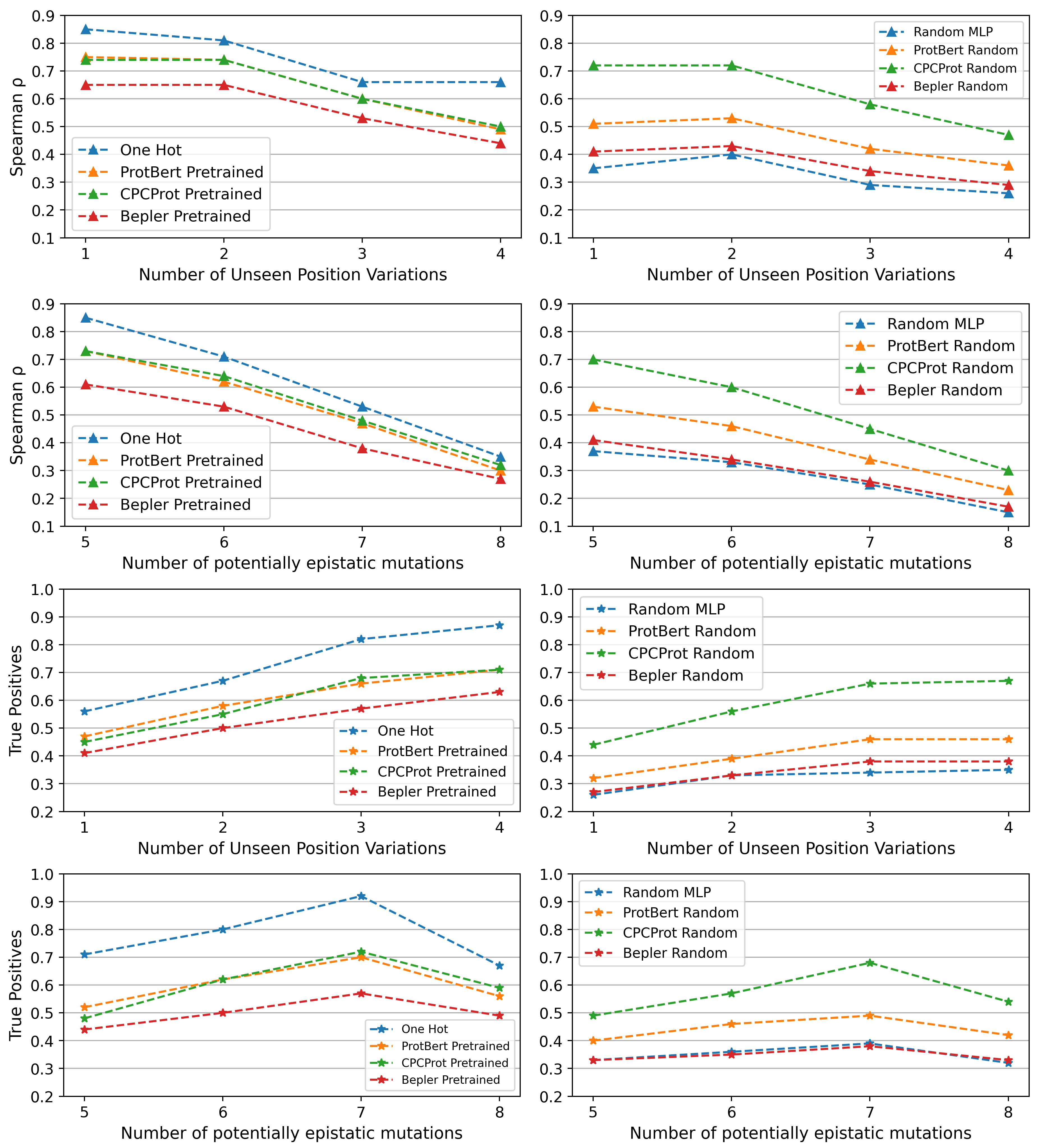}
  \caption{\textbf{Figure 2:} Spearman correlation and true positives metric reported for each test set, all of which are constructed from the GFP dataset \cite{sarkisyan2016local}. The test sets were designed to assess extrapolation outside the training set with the number of unseen position variations and the number of potentially non-linear epistatic mutations. Details on test set composition are provided in the supplemental.}
\end{figure*}
\begin{multicols}{2}
We also assess performance using the "true positives" metric, defined as the fraction of sequences predicted to be in the $80^{\text{th}}$ percentile that are actually in the $80^{\text{th}}$ percentile. This metric is relevant for protein design because it provides an estimate of the expected number of hits out of a set of sequences predicted to have high function. We find that a model having high Spearman correlation does not imply it also having a high true positives score (\textbf{Figure 2}). We find that the one-hot encoding baseline consistently outperforms all pretrained embeddings assessed using both metrics: Spearman's correlation and true positives. 

Interestingly, randomly initialized CPCProt consistently achieve the best test performance compared to other random embedders, performing almost as well as its pretrained counterpart. To further explore this, we used a Random MLP with an embedding dimension of $1024$ which has no architectural prior. We found that the Random MLP consistently performed worse than other random embedders (\textbf{Figure 2}). We speculate that this may be due to the unique inductive bias in CPCProt encoded in its architecture. CPCProt divides a sequence into segments and encodes each sequence segment using a series of convolutional filters \cite{lu2020self}. The outputs of the convolutions are then fed into an autoregressive network. ProtBert and Bepler embedders use bidirectional attention and LSTMs respectively, taking the amino acid at each position as input directly \cite{elnaggar2020prottrans, bepler2019learning}. They do not explicitly extract local sequence patterns (analogous to sequence motifs) which the CNN filters in CPCProt can extract. A similar hypothesis to explain why random embedders with character-level convolution can be stronger baselines than other random embedders for NLP tasks is given in \cite{wang2018can}.

\end{multicols}
\begin{table}[h]
    \centering
    \begin{tabular}{cccccccccccc}
    \hline
    \textbf{Embedding Dimension} & \textbf{2} & \textbf{4} & \textbf{8} & \textbf{16} & \textbf{32} & \textbf{64} & \textbf{128} & \textbf{256} & \textbf{512} & \textbf{1024} & \textbf{2048} \\
    \textbf{GFP2} & -0.01 & 0.02 & 0.04 & 0.04 & 0.06 & 0.09 & 0.09 & 0.10 & 0.10 & 0.11 & 0.11 \\
     \textbf{RM1} & 0.25 & 0.65 & 0.70 & 0.75 & 0.75 & 0.78 & 0.76 & 0.79 & 0.77 & 0.77 & 0.79 \\
     \textbf{RM2} & -0.23 & -0.09 & -0.03 & 0.16 & -0.11 & -0.07 & 0.17 & 0.33 & 0.48 & 0.48 & 0.47 \\
     \textbf{RS1} & 0.03 & 0.64 & 0.45 & 0.70 & 0.62 & 0.74 & 0.69 & 0.66 & 0.68 & 0.69 & 0.72 \\
     \textbf{RS2} & -0.02 & 0.38 & 0.25 & 0.39 & 0.43 & 0.45 & 0.50 & 0.56 & 0.56 & 0.57 & 0.58 \\ 
    \hline
    &&&&&&&&&&&
    \end{tabular}
    \caption{\textbf{Table 2}:  The average Spearman correlation across five initializations of the Random MLP per embedding dimension (\textbf{GFP2}, \textbf{RM1}, \textbf{RM2}) and five random train/test splits with a fixed Random MLP initialization per embedding dimension (\textbf{RS1} and \textbf{RS2}). 1 and 2 refer to the PETase and Rhodopsin datasets described in the supplemental.}
\end{table}
\begin{multicols}{2}

\subsection*{Random Embeddings Perform Better in Higher Dimensions}

Could the differences in performance be also attributable to different embedding dimensions? We test this by observing how varying the embedding dimension of the Random MLP affects prediction performance. We distinguish between two sources of randomness: random embedders and random train/test splits. We find that test performance improves as the random embedding dimension increases, consistent with Figure 1 of \cite{wieting2019no} and Cover's Theorem (\textbf{Table 2}).

\section*{Discussion and Future Work}

Here we showed that one-hot encoding and random embeddings are simple and robust baselines for evaluating embeddings on protein function prediction. In addition to the computational cost of pretraining embedders, methods for interpreting large embedders with tens to hundreds of millions of parameters is not as straightforward as reading off the coefficients in a linear regression model trained using one-hot encoding. The strong performance of randomly initialized CPCProt over other embedders show that model architecture can directly affect embedding quality, even without pretraining. Random embedders that consistently outperform generic encoders with no architectural priors may have inductive biases more suitable for protein function prediction. Thus, random embeddings could be a powerful tool to rapidly search over the space of suitable neural network architectures for protein sequence modelling \cite{he2016powerful, saxe2011random}. 

The consistently strong performance of one-hot encoding suggests that position-wise information is crucial for function prediction. Averaging position-wise embeddings only captures the frequencies of each embedding dimension and destroys local sequence patterns which may be important for function prediction, consistent with the poor performance of averaged one-hot encodings in TAPE \cite{rao2019evaluating}. Methods such as Soft Symmetric Alignment \cite{bepler2019learning}, concatenating all position-wise embeddings, learning a linear combination of position-wise embeddings, or using a single attention layer with one learnable key as the top model may improve embedding performance for protein function prediction.

The observation that random embeddings work at all follows from Cover's Theorem \cite{cover1965geometrical}. Larger embedding dimensions give combinatorially many possible inputs from which linear regression can choose linear combinations \cite{schmidt1992feed}. This means an embedder's performance is partly due to the number of embedding dimensions and may not be entirely attributable to an ability to capture biophysical properties of proteins. This raises concern at methods which aim to interpret model parameters by observing its correlations with some protein property. A random model with enough parameters may, by chance, have parameters that correlate with some property of interest. Thus, random baselines are crucial prerequisites for making conclusions about the effects of pretraining. For representation learning in general, the quality of a representation can only be judged relative to simple, untrained baseline representations.

\subsection*{Acknowledgements}

The authors would like to thank all members of the Moses Lab, Amy Lu, and Tristan Bepler for comments.

\subsection*{Code Availability}

The code to reproduce all the results is available here: \href{http://bit.ly/Random-Embeddings}{http://bit.ly/Random-Embeddings}

\end{multicols}

\newpage
\section*{Supplemental}

\subsection*{DeepSequence}
We use the datasets IF1 \citeSupp{if1}, Ubiquitin (Bolon) \citeSupp{ubiquitin}, BRCA1 (BRCT Domain) \citeSupp{brca1}, $\beta$-Lactamase (Ranganathan) \citeSupp{beta-lac-ranganathan}, Hras \citeSupp{hras}, and $\beta$-Lactamase (Palzkill) \citeSupp{palzkill-2012}. The fitness of each mutation was taken to be the second column of the DeepSequence data \cite{riesselman2018deep}. The top model was trained to predict the raw values without normalization.

\subsection*{GFP Data Splits}
All splits were done on the Sarkisyan et al. GFP dataset \cite{sarkisyan2016local}. The results for GFP2 and GFP3 for each embedder are shown in \textbf{Figure 1}. The results for T2 and T3 are shown in rows 1-2 and 3-4 of \textbf{Figure 2}. Since the data splits are predefined, no random splits are done.
    
    \textbf{GFP2}: Train: training set defined by \cite{rao2019evaluating} with log fluorescence 2.5 or greater. Test: test set defined by \cite{rao2019evaluating} with log fluorescence 2.5 or greater. 
    
    \textbf{GFP3}: Train: log fluorescence within 2.5 to 3.5. Test: log fluorescence above 3.5. Thus, accuracy is assessed on sequences with function values beyond the range observed in the training set.
    
    \textbf{T2}: Train: iterate through the GFP sequences and record the residue indices where mutations exist. Stop until the number of varied residue positions is 120. The training set is constructed from sequences that only vary in these positions. We construct four test sets where there are 1, 2, 3, or 4 positions whose variation was not observed in the training set. The goal is to assess model generalization to unseen residue position variations. \citeSupp{mater2020nk} refers to this as assessing extrapolation performance.
    
    \textbf{T3}: Train: training set defined by \cite{rao2019evaluating}. We stratify the test sets using the number of possibly epistatic mutations in the sequences of the test set defined by \cite{rao2019evaluating}. Two mutations are considered possibly epistatic if both residues are different from wildtype and whose $\beta$ carbons are within 6 Angstroms apart. Hydrogen is taken in place of the $\beta$ carbon for glycines. We construct four test sets where the number possibly epistatic mutations is 5, 6, 7, or 8. The structure 1GFL was used to construct T3 \citeSupp{Yang1997-cv}.
    
\subsection*{Additional Datasets}    
PETase (n=212, train=160) and Rhodopsin (n=798, train=677). The PETase dataset was manually curated from the literature \citeSupp{Austin2018-nb, Cui2021-tf, Han2017-ec, Liu2018-cd, Liu2019-fh, Ma2018-vt, Son2019-fk, Taniguchi2019-zx, Yoshida2016-wr, joo2018structural}. The Rhodopsin dataset was obtained from \citeSupp{karasuyama2018understanding}. For sequences with unequal lengths, shorter sequences were padded with the gap character \texttt{-} on the right to reach the maximum sequence length. For PETase, we predict relative catalytic activity to wildtype. For Rhodopsin, we predict $\lambda_{\text{max}}$. Values to predict were normalized within $[-0.5, 0.5]$.

\newpage
\bibliographystyle{abbrv}
\bibliography{references}

\newpage

\bibliographystyleSupp{abbrv}
\bibliographySupp{supp}

\end{document}